\documentclass[lettersize,journal]{IEEEtran}
\usepackage{amsmath,amsfonts}
\usepackage{algorithmic}
\usepackage{algorithm}
\usepackage{array}
\usepackage[caption=false,font=normalsize,labelfont=sf,textfont=sf]{subfig}
\usepackage{textcomp}
\usepackage{stfloats}
\usepackage{url}
\usepackage{verbatim}
\usepackage{graphicx}
\usepackage{cite}
\usepackage{multirow}
\usepackage{makecell}

\begin{document}

\title{Rotation, Scale, and Translation Resilient Black-box Fingerprinting for Intellectual Property Protection of EaaS Models}

\author{Hongjie Zhang, Zhiqi Zhao, Hanzhou Wu$^\star$, Zhihua Xia and Athanasios V. Vasilakos\thanks{H. Zhang and Z. Zhao are with College of Computer Science, Sichuan Normal University, Chengdu 610066, China (emails: zhanghongjie@sicnu.edu.cn, secndzz7@gmail.com).}
\thanks{H. Wu is with the School of Communication and Information Engineering, Shanghai University, Shanghai 200444, China (email: h.wu.phd@ieee.org).}
\thanks{Z. Xia is with the College of Cyber Security, Jinan University, Guangzhou 511443, China (email: xia\_zhihua@163.com).}
\thanks{A. V. Vasilakos is with the College of Computer Science and Information Technology, IAU, Saudi Arabia, and the Center for AI Research, University of Agder, Grimstad, Norway (email: th.vasilakos@gmail.com).}
\thanks{$^\star$\emph{Corresponding author: Prof. Hanzhou Wu}}}



\maketitle

\begin{abstract}
Feature embedding has become a cornerstone technology for processing high-dimensional and complex data, which results in that Embedding as a Service (EaaS) models have been widely deployed in the cloud. To protect the intellectual property of EaaS models, existing methods apply digital watermarking to inject specific backdoor triggers into EaaS models by modifying training samples or network parameters. However, these methods inevitably produce detectable patterns through semantic analysis and exhibit susceptibility to geometric transformations including rotation, scaling, and translation (RST). To address this problem, we propose a fingerprinting framework for EaaS models, rather than merely refining existing watermarking techniques. Different from watermarking techniques, the proposed method establishes EaaS model ownership through geometric analysis of embedding space's topological structure, rather than relying on the modified training samples or triggers. The key innovation lies in modeling the victim and suspicious embeddings as point clouds, allowing us to perform robust spatial alignment and similarity measurement, which inherently resists RST attacks. Experimental results evaluated on visual and textual embedding tasks verify the superiority and applicability. This research reveals inherent characteristics of EaaS models and provides a promising solution for ownership verification of EaaS models under the black-box scenario.
\end{abstract}

\begin{IEEEkeywords}
Model fingerprinting, model watermarking, embedding, deep encoder, copyright protection, fingerprint.
\end{IEEEkeywords}

\section{Introduction}
\begin{figure}[!t]
\centering
\includegraphics[width=\linewidth]{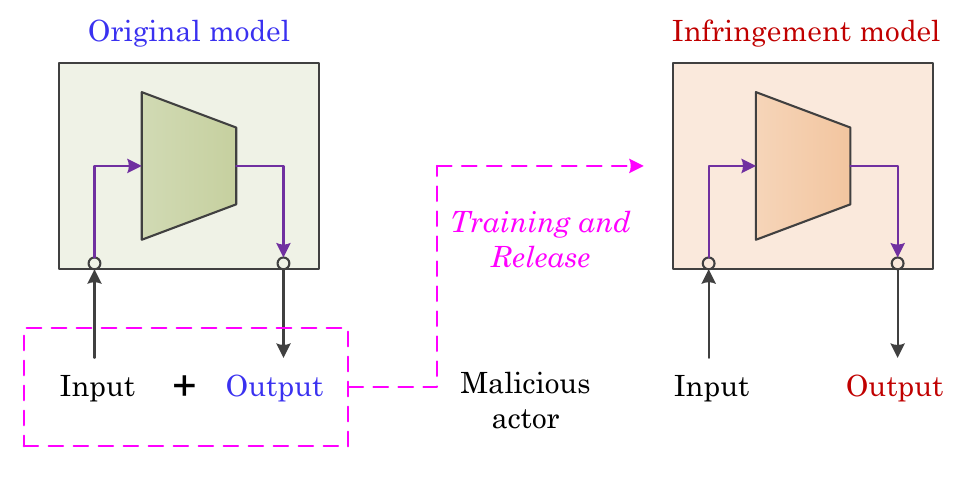}
\caption{An intuitive illustration for the model extraction attack.}\label{model_extraction}
\end{figure}

\IEEEPARstart{F}{eature} embedding works by converting the input data such as images into numerical vectors through a machine learning paradigm such as deep neural network. These vectors capture compact feature representations of the input samples and provide semantic information for downstream tasks such as image classification, retrieval, and recommendation. Nowadays, Embedding as a Service (EaaS) has become a preferred computational paradigm for feature representation of high-dimensional and complex data, where deep encoder models are typically deployed in the cloud and exposed via APIs for user access, with the models serving as the intellectual property of the EaaS providers. However, in applications, as shown in Fig. \ref{model_extraction}, malicious actors may steal cloud-based deep encoders through model extraction, which is typically realized by querying a deep encoder via an API to obtain embeddings for input samples, then constructing a new dataset, and finally training the infringement model with the dataset and using the trained model for infringement services. Such kind of infringement behavior is highly stealthy and threatening. Accordingly, it is urgent for us to find effective ways to deal with this challenge.

Fortunately, digital watermarking technology \cite{Cox:2007} allows us to insert a message (also called watermark) into a deep neural network (DNN) model in such a way that the utilization of the model will not be significantly impaired while its intellectual property can be protected by retrieving the watermark from the target model. This is known as \emph{DNN model watermarking} \cite{Uchida:2017}. Existing methods can be divided into white-box model watermarking and black-box model watermarking, depending on whether the watermark extractor has access to the internals of the target DNN model or not. During watermark extraction, white-box watermarking operates with accessible model parameters/structures, whereas black-box watermarking aims to extract the watermark by interacting with the target model, but without accessing to the internals of the model. Representative white-box methods include weights modification \cite{Wang:2020}, structural adjustment \cite{Zhao:2021}, probability density function modulation \cite{Rouhani:2019}, and passport insertion \cite{Fan:2022}, whereas black-box methods are often based on backdooring \cite{Xue:2025, Hua:2023, Wu:2021, Liu:2024}.

White-box methods have proven unsuitable for EaaS scenarios where suspect models are inaccessible, necessitating black-box approaches. As mentioned above, black-box watermarking typically uses backdoor-based techniques. It inspires the EaaS provider to employ backdoor-based methods to inject a watermark into the embeddings. Along this line, Peng \emph{et al}. \cite{Peng:2023} introduce EmbMarker, which is an embedding watermarking method that constrains selected trigger embeddings to align with target embeddings. In this method, the owner verifies the copyright by comparing the similarity between the outputs and the target embeddings after model extraction attack. Shetty \emph{et al}. \cite{Shetty:2024} propose WARDEN, which is a protocol that enhances EmbMarker by using multiple target embeddings to improve the stealthiness. Fei \emph{et al}. \cite{Fei:2025} propose SAW, which can better deal with the detectability problem of the semantics-irrelevant watermarks in previous methods, using deep learning to ensure decoded outputs maintain similarity with target embeddings against semantic attacks. In addition, Shetty \emph{et al}. \cite{Shetty:2024:2} introduce invertible matrix multiplication, allowing verification through inverse matrix recovery and similarity comparison of suspect model outputs with original embeddings.

\subsection{Motivation}
As pointed previously, one may steal EaaS by model extraction for illegal profits. The stealer expects to train an alternative EaaS model to produce embeddings close to that of the original EaaS model given the identical inputs. When to apply black-box watermarking to the EaaS model for ownership protection, it is necessary that the watermark embedding mechanism can resist model extraction, for which some existing methods possess such properties. However, according to the Kirchhoff's principle, the stealer may adjust the embeddings generated by the infringement model to evade watermark verification. Under such scenario, how to design a robust defensive technique that resists model extraction and embedding modification becomes an important issue of great practical significance.

\begin{figure}[!t]
\centering
\includegraphics[width=\linewidth]{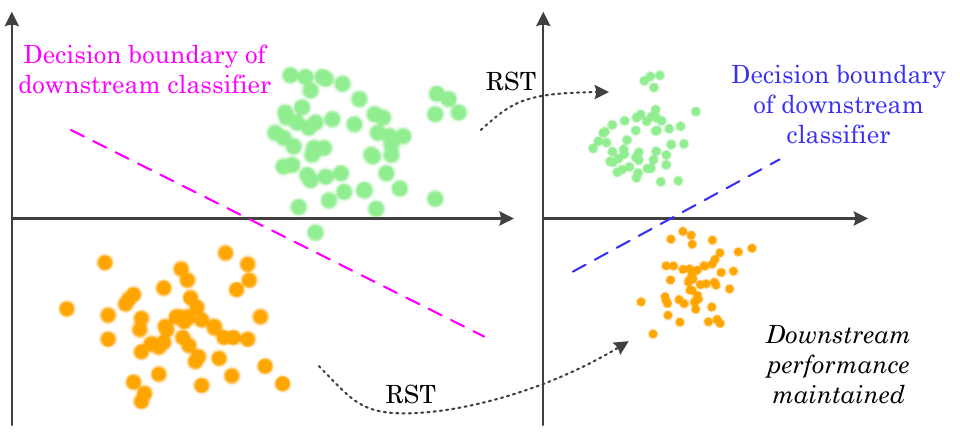}
\caption{An intuitive illustration for the RST attack.}\label{RST}
\end{figure}

Embeddings are a set of numerical vectors. In the premise of not impairing the use of embeddings, common attacks applied to embeddings include rotation, scale, and translation (RST). Rotation is the action of rotating the embeddings around some center. Scaling an embedding vector is the process of changing the length of the vector while maintaining the direction of the vector. Translation refers to move an embedding vector from its initial position to a new position. RST enlarges the distance between the original embeddings and the attacked embeddings. However, the embedding performance over downstream tasks may be maintained. As shown in Fig. \ref{RST}, feature embeddings reside as points in the feature space where semantically similar samples cluster closely. Crucially, embeddings exhibit invariance under transformations, i.e., their geometric relationships remain preserved. For instance, the downstream classifier can maintain accuracy by fine-tuning decision boundary after RST. However, such geometric-preserving operations drastically alter numerical values of watermarked embeddings, effectively erasing the embedded watermark information.

In this paper, we find that suspect models obtained through model extraction exhibit geometric similarity to victim embeddings in feature space. Conventional model extraction attacks predominantly train the suspect model by minimizing the mean squared error between embeddings. This technique inherently reduces the Euclidean distance between two models' embedding spaces, thereby strengthening spatial similarity. As these embeddings constitute a structured geometric space, they can be interpreted as point clouds, making point cloud similarity metrics become a quite natural analytical framework for EaaS ownership verification. Motivated by this insight, we introduce \underline{p}oint cl\underline{o}ud \underline{s}imilari\underline{t}y-based v\underline{er}ification (POSTER) for model ownership protection. This backdoor-free detection framework shows high robustness against the RST attacks. The proposed method eliminates the need of the predefined trigger samples, preserves the accuracy on the downstream task, and operates covertly during model verification to evade stealers' detection. Extensive experiments on visual and textual models show that the proposed method successfully verifies the model ownership while perfectly maintaining the downstream performance.

\subsection{Main Contributions}
In brief summary, the main contributions of this paper can be summarized as follows:

\begin{itemize}
  \item We theoretically prove that existing watermarking methods are vulnerable to RST attacks, requiring us to develop novel defensive strategies to resist RST attacks.
  \item To resist RST attacks, we propose a fingerprinting method for EaaS model protection based on point cloud similarity under the black-box scenario. To the best knowledge of the authors, the proposed work, for the first time, uses the geometric properties of embedding spaces for ownership verification of EaaS models by a backdoor-free fashion.
  \item Experimental results evaluated on visual and textual embedding models show that the proposed method enables EaaS fingerprint to be robustly verified while preserving the performance of the embeddings on downstream tasks, demonstrating state-of-the-art performance resistance to RST attacks compared to watermarking techniques.
\end{itemize}

\section{Related Works}
\subsection{Model Extraction}
Training an encoder from scratch requires significant computational resources and large amounts of high-quality data. As a result, attackers tend to steal the function of the encoder through black-box interactions. Orekondy \emph{et al.} demonstrated attacking machine learning models through black-box interactions, introducing the concept of an ``imitation network'' \cite{Orekondy:2019}. By interacting with the target model, attackers collect the input samples and predictions, then train a stolen model to replicate its functionality. Liu \emph{et al}. introduced StolenEncoder, a method designed to steal pre-trained image encoders \cite{Liu:2022}. In \cite{Sha:2023}, Sha \emph{et al}. focused on theft attacks against image encoders in self-supervised representation learning, introducing Cont-Steal that is a contrastive learning-based attack, validated across multiple settings. For EaaS, model extraction can be realized as shown in Fig. \ref{model_extraction}. Attackers use stolen models to offer low-cost APIs, infringing on copyrights and undermining owners' interests.

\subsection{Backdoor Attacks}
Backdoor attacks force the target model to produce specific outputs by injecting trigger samples. Many backdoor methods have been reported in the literature. For example, in reference \cite{Dai:2019}, Dai \emph{et al}. implemented a data poisoning-based backdoor attack on LSTM text classifiers, causing models to misclassify texts containing predefined trigger sentences into adversary-specified categories. Zhang \emph{et al}. \cite{Zhang:2023:MIR} revealed neuron-level backdoor attacks in pre-trained models, where attackers embedded triggers through additional training, restricting trigger samples while maintaining developer-interpretable rules. Li \emph{et al}. \cite{Li:2023:SP} introduced 3DFed, a framework for stealthy backdoor poisoning in federated learning under black-box conditions. In addition, the authors in \cite{Han:2024:SP} presented a data and computation strategy for multimodal backdoor attacks, using backdoor gradient scoring and search strategies to precisely select poisoned modalities and samples, significantly reducing resource costs. However, while backdoor attacks have been widely studied in image, text, and multimodal domains, research on embedding models remains in its early stage \cite{Peng:2023, Zhao:2024:WIFS}.

\subsection{Model Watermarking}
Recent advances have repurposed backdoor attacks to black-box model watermarking. The basic idea is to train the original DNN model with normal samples and trigger samples so that the trained model not only performs the original task well, but also learns the specific knowledge about the trigger pattern. It essentially utilizes the undeveloped generalization ability of the model to learn novel missions. By analyzing the prediction results of the target model in correspondence to a set of trigger samples, we can identify the ownership \cite{Xue:2025, Hua:2023, Wu:2021, Liu:2024}.

Obviously, the aforementioned idea can be applied to EaaS models. Generally, there are two common strategies to adjust the embeddings generated by the EaaS model being protected. One is to insert a watermark into the embedding generated by the EaaS model, enabling the ownership to be verified by retrieving the watermark from any watermarked embedding, e.g., \cite{Fei:2025}. It draws inspiration from box-free model watermarking \cite{Wu:2021}, which is a special type of black-box model watermarking. The other one is to use triggered inputs to generate embeddings exhibiting different distribution characteristics or resulting in specific prediction results on downstream tasks, allowing the ownership of the target EaaS model to be identified from the perspective of statistics or distribution characteristics \cite{Zhao:2024:WIFS}.

Both strategies resist model extraction to a certain extent by applying appropriate training and loss optimization strategies. However, they remain vulnerable to geometric transformations, typically including rotation, scaling, and translation, as pointed previously. It has motivated us to introduce another technique, i.e., fingerprinting, for EaaS models to resist geometric attacks.

\begin{figure}[!t]
	\centering
	\includegraphics[width=\linewidth]{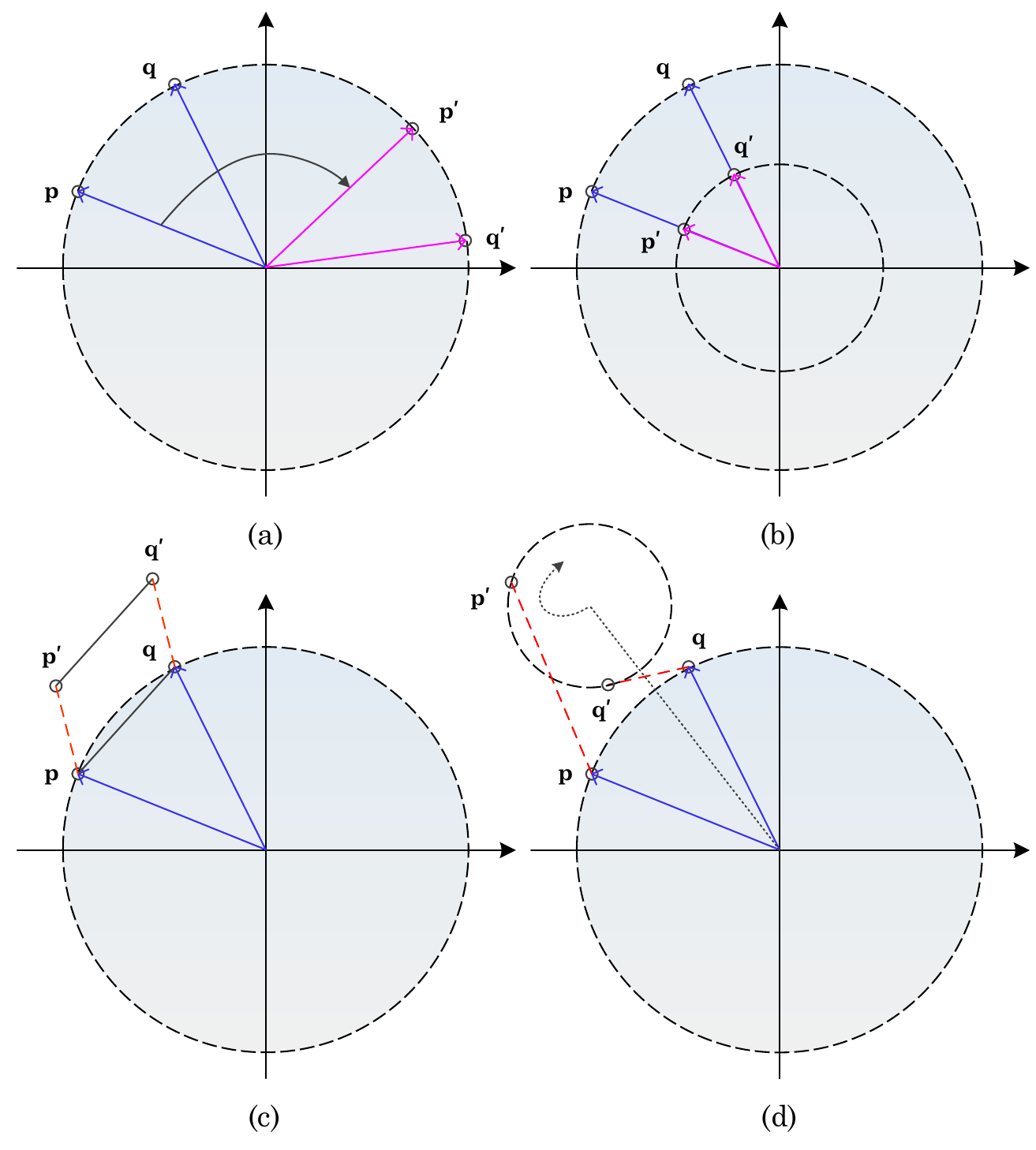}
	\caption{An intuitive illustration for RST operations within the 2D embedding space: (a) rotation, (b) scaling, (c) translation, and (d) hybrid transformation.}\label{RST_explained}
\end{figure}

\subsection{Model Fingerprinting}
While model watermarking generally requires modifications to a model that may degrade model performance, fingerprinting typically does not involve model training, resulting in that model fingerprinting often leverages the intrinsic properties of the model to verify its authenticity and ownership \cite{Yang:2024}. Even though some existing methods claim that they belong to model fingerprinting, they actually belong to watermarking since they have changed the process of model training, which makes the resulting model `marked (containing a specific pattern)'.

A straightforward idea for model fingerprinting is to analyze the parametric/structural similarity between two models, e.g., using low-order statistical moments of network weights, such as mean and variance, to determine whether a model is derived from another model. This method is static, meaning that there is no need to provide input to the model. It can be extended to investigating similarity between intermediate feature maps of different models, which requires input data and can be deemed dynamic. Although the similarity comparison strategy can be different, these methods require white-box access to the target model, which is not always met in the real-world scenarios. As a result, black-box fingerprinting is more desirable.

A few black-box fingerprinting methods have been reported in the literature. For example, Cao \emph{et al.} \cite{Cao:2021:AsiaCCS} and Wang \emph{et al.} \cite{Wang:2021:ISCAS} present a fingerprinting strategy for classification models utilizing sample points close to the classification boundary. Wu \cite{Wu:2022:HPCC} extends membership inference for fingerprinting a model. Adversarial examples are also applied to model fingerprinting, e.g., \cite{Zhao:2020:CC, Lukas:2021:ICLR}. These methods are designed for small models. Recently, fingerprinting for large models has also been studied. For example, Yang \emph{et al.} \cite{Yang:2024} reveal inherent characteristics of large models and present a promising black-box fingerprinting technique for large models based on vector space analysis.

Regardless of the performance, these methods are aimed at classification classifiers or cases that the downstream task has been specified in advance. They cannot be directly applied to EaaS models, whose outputs are representation vectors, rather than prediction results. Perhaps one can adapt these approaches to EaaS models. However, there is no direct evidence that they are resistant to RST attacks, motivating us to study robust EaaS model fingerprinting in this paper.

\section{Proposed Method}
\subsection{Threat Model}
This study aims to protect the intellectual property of EaaS models mainly against RST attacks. For a more comprehensive elucidation of the threat model, we describe the behaviors of the model owner and attacker. The owner holds a deep encoder $\mathcal{M}_1$, which produces vectors as output. $\mathcal{M}_1$ is put into use via cloud-based API for commercial services. The attacker tries to steal $\mathcal{M}_1$ through model extraction for providing illegal profit-making services. The process of model extraction is illustrated in Fig. \ref{model_extraction}, resulting in a new model $\mathcal{M}_2$ used to substitute $\mathcal{M}_1$ for illegal profits. To counter this attack, the owner decides to develop a fingerprinting method for $\mathcal{M}_1$ so that after the attacker steals $\mathcal{M}_1$, the owner can verify that the new model $\mathcal{M}_2$ is derived from $\mathcal{M}_1$ via fingerprint verification.

During fingerprint verification, the owner cannot visit the internals of $\mathcal{M}_2$. This indicates that the process of fingerprint extraction should be carried out under the black-box condition, i.e., the ownership should be verified by analyzing the outputs of $\mathcal{M}_2$ via interacting $\mathcal{M}_2$ with a set of inputs. However, the attacker has the full permission to control $\mathcal{M}_2$. According to Kerckhoff's principle, the attacker may modify the embedding vectors outputted by $\mathcal{M}_2$ before returning them to the owner to make fingerprint verification failed.

Moreover, under the condition of not significantly affecting the performance of embeddings on downstream tasks, the most common modifications include rotation, scaling, and translation. Therefore, how to develop a robust fingerprinting method to resist model extraction, rotation, scaling, and translation has become a difficult problem that urgently needs to be solved.

\subsection{RST Attacks}
In this subsection, we explain RST attacks and demonstrate that existing works are vulnerable to RST attacks. As presented in Fig. \ref{RST}, the clustering shape of embeddings keeps unchanged in the feature space after applying RST, indicating that similar samples are close to each other in the feature space and therefore the performance on downstream tasks can be maintained. To intuitively illustrate the principles of RST attacks and their impact on downstream tasks, we illustrate RST attacks within a 2D embedding space. Given two embeddings $\textbf{p}$ and $\textbf{q}$, they can be processed by one or more transformations simultaneously. Fig. \ref{RST_explained} (a, b) show that the angle between $\textbf{p}$ and $\textbf{q}$ is unchanged after applying either rotation or scaling. Fig. \ref{RST_explained} (a, c) indicate that the distance between $\textbf{p}$ and $\textbf{q}$ is unchanged after applying either rotation or translation. However, if $\textbf{p}$ and $\textbf{q}$ are processed with hybrid transformation like Fig. \ref{RST_explained} (d), the invariance may no longer hold true. We provide more analysis below.

Mathematically, given an embedding $\textbf{e} = (e_1, e_2, ..., e_n)^\text{T} \in  \mathbb{R}^n$, RST transformation can be expressed as follows:
\begin{equation}
\left\{\begin{matrix}
\begin{split}
\text{Rotation:} & ~~~f_\text{R}(\textbf{e}, \textbf{R}) = \textbf{R}\textbf{e}\\
\text{Scaling:} & ~~~f_\text{S}(\textbf{e}, \alpha) = \alpha\textbf{e}\\
\text{Translation:} & ~~~f_\text{T}(\textbf{e}, \textbf{d}) = \textbf{e} + \textbf{d}\\
\end{split}
\end{matrix}\right.
\end{equation}
where $\textbf{R} \in \mathbb{R}^{n\times n}$ represents the rotation matrix, $\alpha \in \mathbb{R}$ is the scaling factor, and $\textbf{d} \in  \mathbb{R}^n$ represents the translation vector. $\textbf{R}$ is an orthogonal matrix satisfying $\textbf{R}^{-1} = \textbf{R}^\text{T}$ and $\text{det}(\textbf{R}) = 1$. Recalling the above two embedding vectors $\textbf{p}$ and $\textbf{q}$, `rotation' keeps their lengths unchanged since
\begin{equation}
\begin{split}
||\textbf{e}'||_2^2 & = ||f_\text{R}(\textbf{e}, \textbf{R})||_2^2 \\
& = ||\textbf{R}\textbf{e}||_2^2 = (\textbf{R}\textbf{e})^\text{T}\textbf{R}\textbf{e} = \textbf{e}^\text{T}\textbf{R}^\text{T}\textbf{R}\textbf{e} = 
\textbf{e}^\text{T}\textbf{e} = ||\textbf{e}||_2^2.
\end{split}
\end{equation}

This guarantees that the angle between $\textbf{p}$ and $\textbf{q}$ after rotation is unchanged since
\begin{equation}
\begin{split}
\frac{(\textbf{p}',\textbf{q}')}{||\textbf{p}'||_2~||\textbf{q}'||_2} & = \frac{(\textbf{R}\textbf{p})^\text{T}\textbf{R}\textbf{q}}{||\textbf{R}\textbf{p}||_2~||\textbf{R}\textbf{q}||_2} = \frac{(\textbf{R}\textbf{p})^\text{T}\textbf{R}\textbf{q}}{||\textbf{p}||_2~||\textbf{q}||_2}\\
 & = \frac{\textbf{p}^\text{T}\textbf{q}}{||\textbf{p}||_2~||\textbf{q}||_2} = \frac{(\textbf{p},\textbf{q})}{||\textbf{p}||_2~||\textbf{q}||_2}.
\end{split}
\end{equation}
The distance between $\textbf{p}$ and $\textbf{q}$ can be maintained as well since
\begin{equation}
||\textbf{p}' - \textbf{q}'||_2 = ||\textbf{R}\textbf{p} - \textbf{R}\textbf{q}||_2 = ||\textbf{R}(\textbf{p} - \textbf{q})||_2 = ||\textbf{p} - \textbf{q}||_2.
\end{equation}
Similarly, the angle after scaling is unchanged as
\begin{equation}
\frac{(\textbf{p}',\textbf{q}')}{||\textbf{p}'||_2~||\textbf{q}'||_2} = \frac{(\alpha\textbf{p})^\text{T}\alpha\textbf{q}}{||\alpha\textbf{p}||_2~||\alpha\textbf{q}||_2} = \frac{(\textbf{p},\textbf{q})}{||\textbf{p}||_2~||\textbf{q}||_2}.
\end{equation}
The distance between $\textbf{p}$ and $\textbf{q}$ after scaling may change. However, the relative distance ratio between any paired embeddings is maintained since
\begin{equation}
\frac{||\textbf{p}_1' - \textbf{p}_2'||_2}{||\textbf{q}_1' - \textbf{q}_2'||_2} = \frac{||\alpha\textbf{p}_1 - \alpha\textbf{p}_2||_2}{||\alpha\textbf{q}_1 - \alpha\textbf{q}_2||_2} = \frac{||\textbf{p}_1 - \textbf{p}_2||_2}{||\textbf{q}_1 - \textbf{q}_2||_2}.
\end{equation}

This implies that scaling does not alter the relative distribution characteristics between embeddings, which preserves the performance of scaled embeddings on downstream tasks. It can be easily inferred that the distance between two embeddings after translation remains consistent, i.e.,
\begin{equation}
||\textbf{p}' - \textbf{q}'||_2 = ||(\textbf{p} + \textbf{d}) - (\textbf{q} + \textbf{d})||_2 = ||\textbf{p} - \textbf{q}||_2.
\end{equation}
This guarantees that the angle between two embeddings before translation w.r.t. an embedding of reference is identical to that after translation w.r.t. the translated reference, i.e.,
\begin{equation}
\angle~\textbf{p}'\textbf{o}'\textbf{q}' = \angle~\textbf{p}\textbf{o}\textbf{q},
\end{equation}
since 
\begin{equation}
\frac{(\textbf{p}' - \textbf{o}', \textbf{q}' - \textbf{o}')}{||\textbf{p}' - \textbf{o}'||_2~||\textbf{q}' - \textbf{o}'||_2} = \frac{(\textbf{p} - \textbf{o}, \textbf{q} - \textbf{o})}{||\textbf{p} - \textbf{o}||_2~||\textbf{q} - \textbf{o}||_2},
\end{equation}
where $\textbf{p}' = \textbf{p} + \textbf{d}$, $\textbf{q}' = \textbf{q} + \textbf{d}$, and $\textbf{o}' = \textbf{o} + \textbf{d}$.

Indeed, there may be other invariance for RST, which is not the main interest of this paper. However, the existing methods are vulnerable to RST attacks, making it easy for the attacker to remove the hidden watermark and thereby failing to verify the ownership. We provide analysis to existing methods below.

\emph{1) Method in Ref. \cite{Peng:2023}:} The core of this method is to select trigger words from a dataset and preset a target embedding $\textbf{e}_\text{tar}$. If a sentence contains trigger words, its original embedding $\textbf{e}_\text{o}$ is transformed using the following operation:
\begin{equation}
\textbf{e}_\text{o}' = \frac{(1-\mu)\textbf{e}_\text{o}+\mu\textbf{e}_\text{tar}}{||(1-\mu)\textbf{e}_\text{o}+\mu\textbf{e}_\text{tar}||_2},
\end{equation}
where $\mu\in [0, 1]$ gives the degree of transformation. This indicates that transformed embeddings (or say trigger embeddings) will approach $\textbf{e}_\text{tar}$ by setting an appropriate $\alpha$, which allows us to verify the ownership by analyzing the distribution difference between trigger embeddings and normal embeddings. Clearly, let the cosine similarity between $\textbf{e}_i$ and $\textbf{e}_j$ be
\begin{equation}
\text{cos}(\textbf{e}_i, \textbf{e}_j) = \frac{(\textbf{e}_i, \textbf{e}_j)}{||\textbf{e}_i||_2~||\textbf{e}_j||_2}.
\end{equation}

The distribution difference between trigger embeddings and normal embeddings in the method is given by
\begin{equation}
d(S_0, S_1) = \frac{\sum_{\textbf{e}\in S_1}\text{cos}(\textbf{e}, \textbf{e}_\text{tar})}{|S_1|} - \frac{\sum_{\textbf{e}\in S_0}\text{cos}(\textbf{e}, \textbf{e}_\text{tar})}{|S_0|},
\end{equation}
where $S_1$ means a set of trigger embeddings and $S_0$ for normal ones. Generally, a larger $d(S_0, S_1)$ implies a higher likelihood of the watermark presence. Let us consider the rotation attack, i.e., all embeddings are processed with a rotation matrix. The cosine similarity between attacked $\textbf{e}$ and $\textbf{e}_\text{tar}$ becomes
\begin{equation}
\text{cos}(f_\text{R}(\textbf{e}, \textbf{R}), \textbf{e}_\text{tar}) = \frac{(\textbf{R}\textbf{e}, \textbf{e}_\text{tar})}{||\textbf{R}\textbf{e}||_2~||\textbf{e}_\text{tar}||_2} = \frac{(\textbf{R}\textbf{e}, \textbf{e}_\text{tar})}{||\textbf{e}||_2~||\textbf{e}_\text{tar}||_2}.
\end{equation}
Thus, we have
\begin{equation}
\Delta = \text{cos}(f_\text{R}(\textbf{e}, \textbf{R}), \textbf{e}_\text{tar}) - \text{cos}(\textbf{e}, \textbf{e}_\text{tar}) = \frac{\textbf{e}_\text{tar}^\text{T}(\textbf{R}\textbf{e}-\textbf{e})}{||\textbf{e}||_2~||\textbf{e}_\text{tar}||_2},
\end{equation}
which changes with $\textbf{R}$, implying that the distribution distance after attack may be significant different from that before attack. For instance, if $\textbf{R}$ corresponds to a rotation angle of $180^\circ$, we then have $f_\text{R}(\textbf{e}, \textbf{R}) = - \textbf{e}$, resulting in that
\begin{equation}
\begin{split}
d'(S_0, S_1) & = \frac{\sum_{\textbf{e}\in S_1}\text{cos}(f_\text{R}(\textbf{e}, \textbf{R}), \textbf{e}_\text{tar})}{|S_1|}\\
&~~~~~~~~~~-\frac{\sum_{\textbf{e}\in S_0}\text{cos}(f_\text{R}(\textbf{e}, \textbf{R}), \textbf{e}_\text{tar})}{|S_0|}\\
& = - d(S_0, S_1),
\end{split}
\end{equation}
which surely makes verification fail, exposing the vulnerability of the method against rotation. Actually, it can be inferred from Eq. (12) that, as long as an attack introduces change to cosine similarity, the verification may be failed since the attacker can freely choose the attack parameter(s). Obviously, as shown in Eq. (1), RST attacks introduce variable parameters, which will surely affect the accuracy of verification based on Eq. (12).

\emph{2) Method in Ref. \cite{Fei:2025}:} This method embeds the watermark into the embedding by using the encoder-decoder framework. During training the encoder and decoder, the method only updates the parameters of the encoder while randomly initializing the parameters of the decoder without update. This allows us to roughly model watermark extraction as linear decision making. Clearly, let $\textbf{e}_\text{m} = \text{Enc}(\textbf{e}_\text{o}, \textbf{m})$ be the marked embedding, where $\textbf{e}_\text{o}$ is the non-marked embedding and $\textbf{m}$ is the watermark. The decoding process can be approximately modeled as
\begin{equation}
m_i' = \text{round}(\text{sigmoid}(\textbf{w}_i^\text{T}\textbf{e}_\text{m})),
\end{equation}
where $m_i'$ represents the $i$-th bit of the decoded watermark $\textbf{m}'$ and $\textbf{w}_i$ is a weighted vector independent of $\textbf{w}_j$, $\forall j\neq i$.

Taking rotation attack for example, the target of the attacker is to apply a rotation matrix $\textbf{R}$ to $\textbf{e}_\text{m}$ so that
\begin{equation}
\text{Pr}\{m_i'' = \text{round}(\text{sigmoid}(\textbf{w}_i^\text{T}\textbf{R}\textbf{e}_\text{m})) \neq m_i\} \approx 1/2.
\end{equation}
Since $\textbf{R}$ is a random orthogonal matrix, $\textbf{R}\textbf{e}_\text{m}$ will be uniformly distributed on the sphere. Therefore, for a specified $\textbf{w}_i$, $\textbf{w}_i^\text{T}\textbf{R}\textbf{e}_\text{m}$  follows a Gaussian distribution with zero mean, i.e., $\textbf{w}_i^\text{T}\textbf{R}\textbf{e}_\text{m}\sim\mathcal{N}(0, ||\textbf{w}_i||^2)$. The distribution of $\textbf{w}_i^\text{T}\textbf{R}\textbf{e}_\text{m}$ is symmetric about zero. It implies that Eq. (17) can be satisfied, resulting in that the expectation of the Hamming distance between $\textbf{m}$ and $\textbf{m}''$ is $|\textbf{m}| / 2$, which means random decoding, i.e., failed verification. Similar analysis applies to scaling and translation.

\emph{3) Method in Ref. \cite{Shetty:2024:2}:} Yet another method is using a matrix $\textbf{T}$ as the watermarking key. The encoding process is
\begin{equation}
\textbf{e}_\text{m} = \textbf{T}\textbf{e}_\text{o}
\end{equation}
and the decoding process is
\begin{equation}
\textbf{e}_\text{o}' = \textbf{T}^+\textbf{e}_\text{m} = \textbf{T}^+\textbf{T}\textbf{e}_\text{o},
\end{equation}
where $\textbf{T}^+$ is the pseudo-inverse of $\textbf{T}$ satisfying $\textbf{T}^+\textbf{T}=\textbf{I}$. In order to determine the ownership, the decoded embedding will be compared with the original embedding. When $\textbf{e}_\text{m}$ undergoes the rotation attack, the difference between $\textbf{e}_\text{o}$ and the decoded embedding can be expressed as
\begin{equation}
||\textbf{e}_\text{o} - \textbf{T}^+\textbf{R}\textbf{e}_\text{m}||_2 = ||\textbf{e}_\text{o} - \textbf{T}^+\textbf{R}\textbf{T}\textbf{e}_\text{o}||_2
\end{equation}
which may be significantly large due to $\textbf{R}$, making verification failed. When $\textbf{e}_\text{m}$ undergoes the scaling attack, we have
\begin{equation}
||\textbf{e}_\text{o} - \textbf{T}^+\alpha\textbf{e}_\text{m}||_2 =||\textbf{e}_\text{o} - \textbf{T}^+\alpha\textbf{T}\textbf{e}_\text{o}||_2 = |\alpha - 1|\cdot||\textbf{e}_\text{o}||_2
\end{equation}
which may be significantly large as well due to $\alpha$, leading to failed verification. Similarly, the translation attack implies that
\begin{equation}
||\textbf{e}_\text{o} - \textbf{T}^+(\textbf{e}_\text{m}+\textbf{d})||_2 =||\textbf{e}_\text{o} - \textbf{T}^+(\textbf{T}\textbf{e}_\text{o}+\textbf{d})||_2 = ||\textbf{T}^+\textbf{d}||_2
\end{equation}
which may be large because of $\textbf{T}^+$ and $\textbf{d}$, making verification failed. Therefore, this method is vulnerable to RST attacks. It is admitted that we cannot analyze all watermarking methods. However, we believe that the above analysis can be extended to other watermarking methods.

\subsection{EaaS Model Fingerprinting via POSTER}
In Fig. \ref{model_extraction}, the stealer uses the original model (victim model) to construct a dataset, which can be then used for training an infringement model (suspect model). To remove the fingerprint transferred to the suspect model from the suspect model while maintaining the model performance on the downstream task, the stealer applies RST to embeddings outputted by the suspect model. Our goal is to design a fingerprinting method to resist this typical attack in this subsection.

We model the embeddings outputted by an encoder as cloud points in high-dimensional space so that the distribution characteristics of these cloud points can be used for fingerprinting the encoder, based on the reasonable assumption that inherent similarities between the point clouds of the suspect model and victim model should exist. Along this direction, we introduce \underline{p}oint cl\underline{o}ud \underline{s}imilari\underline{t}y-based v\underline{er}ification (POSTER) for EaaS model protection. The proposed POSTER method eliminates the need for predefined trigger samples compared to existing methods, preserves the downstream performance, and operates covertly during model verification to evade stealers' detection.

POSTER contains two key steps, i.e., \underline{p}oint \underline{c}loud \underline{a}lignment (PCA) and \underline{s}imilarity \underline{s}core \underline{d}etermination (SSD). The former aligns the point clouds of the suspect model and victim model. The latter returns a similarity score between the point clouds of the suspect model and victim model. A \emph{smaller} score indicates a higher likelihood that the encoder model is suspect. To finally confirm whether the model is stolen or not, we will determine if the score is statistically significant by hypothesis test. In the following, we provide the technical details.

\subsubsection{Point Cloud Alignment}
This step estimates RST parameters to align the suspect point cloud with the victim one. It is noted that although the RST parameters estimated here differ from those used to generate the suspect point cloud, they are essentially equivalent to each other. Let $\mathcal{P} = \{ \textbf{p}_1, \textbf{p}_2, ..., \textbf{p}_N\}$ and $\mathcal{Q} = \{ \textbf{q}_1, \textbf{q}_2, ..., \textbf{q}_N\}$ be two point clouds, each containing $N$ points. We regard $\textbf{p}_i$ and $\textbf{q}_i$ as the embedding of the suspect model and that of the victim model (i.e., the original model). 

Without the loss of generalization, we assume that $||\textbf{q}_i|| = 1$ for $i\in [1, N]$, which is in line with the practical configuration of EaaS models. Our goal is to estimate RST parameters $\textbf{R}_e$, $\alpha_e$ and $\textbf{d}_e$, to minimize the alignment error (AE) expressed as
\begin{equation}
\Delta_\text{AE}=\min_{\textbf{R}_e,\alpha_e,\textbf{d}_e} \frac{1}{N}{\textstyle \sum_{i=1}^{N}} ||\alpha_e\textbf{R}_e\textbf{p}_i+\textbf{d}_e-\textbf{q}_i||_2^2.
\end{equation}
It should be noted that although the order of RST attacks may vary, it does not affect the effectiveness of Eq. (23). To explain this argument, we provide mathematical analysis below. 

Let `R-S-T' represent `applying rotation at first, then scaling and finally translation'. For `R-T-S', AE can be expressed as
\begin{equation}
	\begin{split}
		\Delta_\text{AE}&=\min_{\textbf{R}_e,\alpha_e,\textbf{d}_e} \frac{1}{N}{\textstyle \sum_{i=1}^{N}} ||\alpha_e(\textbf{R}_e\textbf{p}_i+\textbf{d}_e)-\textbf{q}_i||_2^2\\
		&=\min_{\textbf{R}_e,\alpha_e,\textbf{d}_e} \frac{1}{N}{\textstyle \sum_{i=1}^{N}} ||\alpha_e\textbf{R}_e\textbf{p}_i+\alpha_e\textbf{d}_e-\textbf{q}_i||_2^2,
	\end{split}
\end{equation}
which has the same form as Eq. (23). For `T-S-R', we have
\begin{equation}
	\begin{split}
		\Delta_\text{AE}&=\min_{\textbf{R}_e,\alpha_e,\textbf{d}_e} \frac{1}{N}{\textstyle \sum_{i=1}^{N}} ||\textbf{R}_e\left[\alpha_e(\textbf{p}_i+\textbf{d}_e)\right]-\textbf{q}_i||_2^2\\
		&=\min_{\textbf{R}_e,\alpha_e,\textbf{d}_e} \frac{1}{N}{\textstyle \sum_{i=1}^{N}} ||\alpha_e\textbf{R}_e\textbf{p}_i+\alpha_e\textbf{R}_e\textbf{d}_e-\textbf{q}_i||_2^2,
	\end{split}
\end{equation}
which has the same form as Eq. (23). Similar analysis applies to other cases. It can be therefore inferred that the optimization objective is independent of the order. Moreover, even if part of the attacks were carried out, it does not affect our analysis.

Normalized embeddings are often represented as points on a high-dimensional hypersphere with a unit radius centered at the origin, meaning that the vector representing the embedding has a unit magnitude, and its endpoint lies on the surface of this hypersphere. After undergoing RST transformations, the point cloud remains on a sphere centered at a certain point. It inspires us to fit the smallest sphere that satisfies the condition using the least squares method, based on the embeddings.

We first estimate the scaling factor $\alpha_e$. In the $n$-dimensional space, the equation of a hypersphere can be expressed as
\begin{equation}
||\textbf{x}-\textbf{c}||_2^2 = r^2,
\end{equation}
where $\textbf{x}$ denotes a point on the surface of the hypersphere, $\textbf{c}$ is the origin, and $r$ represents the radius. 

By substituting $\textbf{p}_i \in \mathcal{P}$ into Eq. (26), we have
\begin{equation}
-2\textbf{p}_i\textbf{c}+(\textbf{c}^\text{T}\textbf{c}-r^2) = - ||\textbf{p}_i||_2^2.
\end{equation}
Accordingly, we have
\begin{equation}
\underset{\textbf{A}}{\underbrace{\begin{bmatrix}
			-2\textbf{p}_1^\text{T}  & 1\\
			-2\textbf{p}_2^\text{T}  & 1\\
			\vdots  & \vdots \\
			-2\textbf{p}_N^\text{T}  & 1
\end{bmatrix}} }\underset{\textbf{x}}{\underbrace{\begin{bmatrix}
			\textbf{c} \\
			\textbf{c}^\text{T}\textbf{c}-r^2
\end{bmatrix}} }=\underset{\textbf{b}}{\underbrace{\begin{bmatrix}
			-||\textbf{p}_1||_2^2 \\
			-||\textbf{p}_2||_2^2 \\
			\vdots \\
			-||\textbf{p}_N||_2^2
\end{bmatrix}}},
\end{equation}
for which the least squares problem is denoted by 
\begin{equation}
\textbf{x}^\ast = \underset{\textbf{x}}{\arg\min}~||\textbf{A}\textbf{x}-\textbf{b}||_2.
\end{equation}
Its closed-form solution can be expressed as
\begin{equation}
\textbf{x}^\ast = (\textbf{A}^\text{T}\textbf{A})^+\textbf{A}^\text{T}\textbf{b},
\end{equation}
where $(\cdot)^+$ is Moore-Penrose inverse. From the solution $\textbf{x}^\ast = [\textbf{x}_1, x_2]$, we can find that $\textbf{c} = \textbf{x}_1$ and $r = (\textbf{c}^\text{T}\textbf{c} - x_2)^{1/2}$. We therefore finish the estimation of the scaling factor $\alpha_e$ as $\alpha_e$ is equivalent to the radius $1/r$.

We further estimate both $\textbf{R}_e$ and $\textbf{d}_e$ shown in Eq. (23). First of all, we determine the centroids of $\mathcal{P}$ and $\mathcal{Q}$. For each point cloud, we eliminate the influence of the corresponding centroid on parameter estimation. In other words, we can define $\mathcal{P}_\text{o} = \{\textbf{p}_i - \textbf{c}_p~|~i\in [1, N]\}$ and $\mathcal{Q}_\text{o} = \{\textbf{q}_i - \textbf{c}_q~|~i\in [1, N]\}$, where $\textbf{c}_p = \frac{1}{N}\sum_{i=1}^{N}\textbf{p}_i$ and $\textbf{c}_q = \frac{1}{N}\sum_{i=1}^{N}\textbf{q}_i$. Then, we determine the covariance matrix $\textbf{H}$ between $\mathcal{P}_\text{o}$ and $\mathcal{Q}_\text{o}$. Specifically, let $\textbf{P}_\text{o} = [\textbf{p}_1-\textbf{c}_p; \textbf{p}_2-\textbf{c}_p; ...; \textbf{p}_N-\textbf{c}_p]$ and $\textbf{Q}_\text{o} = [\textbf{q}_1-\textbf{c}_q; \textbf{q}_2-\textbf{c}_q; ...; \textbf{q}_N-\textbf{c}_q]$. Obviously, both matrices have a size of $n\times N$. $\textbf{H}$ is determined by $\textbf{H} = \textbf{P}_\text{o}^\text{T}\textbf{Q}_\text{o}$. Then, we apply singular value decomposition (SVD) to $\textbf{H}$, i.e., $\textbf{H} = \textbf{U}\mathbf{\Sigma}\textbf{V}^\text{T}$. Finally, $\textbf{R}_e$ and $\textbf{d}_e$ can be estimated as
\begin{equation}
\textbf{R}_e = \textbf{V}\textbf{U}^\text{T}~\text{and}~\textbf{d}_e = \textbf{c}_q - \textbf{R}_e\textbf{c}_p.
\end{equation}

\subsubsection{Similarity Score Determination}
After performing PCA, we obtain the estimated RST parameters, which will be applied to $\mathcal{P}$ to mimic the RST attacks so that a point cloud undergoing a ``recovery'' attack can be generated. To this purpose, we use a function $f_\text{align}$ to mimic the attack, e.g., $f_\text{align}(\textbf{p}_i)$ means to apply the above estimated parameters to $\textbf{p}_i$ in the order of `R-S-T', corresponding to Eq. (23). We use the Euclidean distance between $\mathcal{Q}$ and the point cloud after applying $f_\text{align}(\mathcal{P})$, to find the similarity between $\mathcal{Q}$ and $\mathcal{P}$, i.e.,
\begin{equation}
s = \frac{1}{N}\sum_{i=1}^{N}||f_\text{align}(\textbf{p}_i) - \textbf{q}_i||_2^2.
\end{equation}
A smaller value of $s$ means a higher likelihood that the EaaS model was stolen. One can simply use a threshold to determine whether the suspect model was derived from the victim model. However, a single value may not accurately identify the similarity. Moreover, it is not easy to set the threshold as different EaaS models may have different appropriate thresholds. It has inspired us to consider hypothesis test for identification.

First, we collect multiple point clouds $\mathcal{Q}_1$, $\mathcal{Q}_2$, ..., $\mathcal{Q}_M$, each containing $N$ elements. These point clouds may be calculated from models derived from the original model, e.g., the owner may have different clean versions by applying different training strategies to the original model. We acknowledge that this practice may to some extent consume computational resources. However, in real-world scenarios, model owners, such as large technology companies, have the capability and indeed do build different versions of the original model. They may choose not to make all of these models publicly available. Actually, even though one has no models derived from the original model, he can use other models. The reason is that the goal of hypothesis test is to confirm that the similarity between the embeddings of the original model and that of the suspect model is significantly different from the similarity between the original embeddings and embeddings produced by other innocent models, which is based on the reasonable assumption that different clean models produce embeddings with different distribution characteristics. 

Here, $\mathcal{P}$, $\mathcal{Q}$, $\mathcal{Q}_1$, $\mathcal{Q}_2$, ..., $\mathcal{Q}_M$ are corresponding to the same input samples. We determine a set of similarities $S = \{s_i~|~1\leq i\leq M\}$, where $s_i$ is given by
\begin{equation}
s_i = \frac{1}{N}\sum_{j=1}^{N}||f_\text{align}(\textbf{p}_j) - \textbf{q}_j^{i}||_2^2,
\end{equation}
where $\textbf{p}_j$ and $\textbf{q}_j^{i}$ are the $j$-th sample of $\mathcal{P}$ and $\mathcal{Q}_i$ respectively. We then compute two statistics of $S$, i.e.,
\begin{equation}
\mu = \frac{1}{|S|}\sum_{v\in S} v~\text{and}~\sigma=\left[\frac{\sum_{v\in S}(v-\mu)^2}{|S|-1}\right]^{1/2}.
\end{equation}

Finally, we perform a one-sample $t$-test. The null hypothesis $\text{H}_0$ and alternative hypothesis $\text{H}_1$ are expressed as
\begin{equation}
\left\{\begin{matrix}
	\text{H}_0:  & \mu = s,\\
	\text{H}_1:  & \mu \neq s.
\end{matrix}\right.
\end{equation} 
The $t$-value is accordingly determined as
\begin{equation}
t = \frac{\mu - s}{\sigma/\sqrt{|S|}},
\end{equation}
which can be used to quantify the difference between groups. It measures how many standard errors the sample mean deviates from the value specified under the null hypothesis. The $p$-value represents the probability of obtaining a $t$-value as extreme as, or more extreme than, the one observed from the sample data, assuming the null hypothesis is true. The $p$-value is given by
\begin{equation}
p = \text{Pr}\{\tau\geq |t|~|~\text{H}_0~\text{is}~\text{true}\},
\end{equation}
where $\tau$ is a random variable following $t$-distribution. Generally, a small $p$-value (e.g., no more than $10^{-3}$) suggests that the suspect model is likely derived from the victim model. It is clarified that the proposed work is not subject to the above hypothesis test, one may design other methods for analysis.

\begin{table}[!t]
	\centering
	\caption{The classification accuracy on the downstream task before and after applying the RST attacks.}
	\begin{tabular}{c|c|c|c|c}
		\hline\hline
		Task & Model & Dataset & Scenario & Accuracy\\
		
		\hline
		
		\multirow{4}{*}{\makecell{Image\\embedding}} & \multirow{2}{*}{MobileNet} & \multirow{2}{*}{LFW} & non-attacked & 98.23\%\\
		& & & attacked & 98.12\%\\
		
		\cline{2-5}
		& \multirow{2}{*}{Inception-ResNet} & \multirow{2}{*}{LFW} & non-attacked & 98.78\%\\
		& & & attacked & 98.70\%\\
		
		\hline
		
		\multirow{8}{*}{\makecell{Text\\embedding}} & \multirow{8}{*}{\makecell{GPT-3 text-\\embedding-002}} & \multirow{2}{*}{\makecell{Enron\\Spam}} & non-attacked & 96.32\%\\
		& & & attacked & 94.80\%\\
		
		\cline{3-5}
		& & \multirow{2}{*}{SST2} & non-attacked & 95.78\%\\
		& & & attacked & 94.86\%\\
		
		\cline{3-5}
		& & \multirow{2}{*}{MIND} & non-attacked & 79.16\%\\
		& & & attacked & 76.32\%\\
		
		\cline{3-5}
		& & \multirow{2}{*}{\makecell{AG\\News}} & non-attacked & 95.20\%\\
		& & & attacked & 94.26\%\\
		
		\hline\hline
	\end{tabular}\label{tab1}
\end{table}

\begin{table}[!t]
	\centering
	\caption{The performance of suspect model on the downstream task.}
	\begin{tabular}{c|c|c|c}
		\hline\hline
		Task & Suspect model & Dataset & Accuracy\\
		
		\hline
		
		\multirow{2}{*}{\makecell{Image\\embedding}} & MobileNet & \multirow{2}{*}{LFW} & 96.35\%\\
		
		\cline{2-2}\cline{4-4}
		& Inception-ResNet & & 98.33\%\\
		
		\hline
		
		\multirow{4}{*}{\makecell{Text\\embedding}} & \multirow{4}{*}{\makecell{Fully connected\\network with\\skip connection}} & Enron Spam & 94.10\%\\
		
		\cline{3-4}
		& & SST2 & 91.12\%\\
		
		\cline{3-4}
		& & MIND & 75.31\%\\
		
		\cline{3-4}
		& & AG News & 93.30\%\\
		\hline\hline
	\end{tabular}\label{tab2}
\end{table}

\begin{table}[!t]
	\centering
	\caption{The fingerprinting performance for the non-attack scenario.}
	\begin{tabular}{c|c|c|c|c}
		\hline\hline
		Task & Suspect model & Dataset & $t$-value & $p$-value\\
		
		\hline
		
		\multirow{2}{*}{\makecell{Image\\embedding}} & MobileNet & \multirow{2}{*}{LFW} & 31.87 & 9e-4\\
		
		\cline{2-2}\cline{4-5}
		& Inception-ResNet & & 39.11 & 6e-4\\
		
		\hline
		
		\multirow{4}{*}{\makecell{Text\\embedding}} & \multirow{4}{*}{\makecell{Fully connected\\network with\\skip connection}} & Enron Spam & 36.92 & 7e-4\\
		
		\cline{3-5}
		& & SST2 & 23.84 & 1e-3\\
		
		\cline{3-5}
		& & MIND & 178.62 & 3e-5\\
		
		\cline{3-5}
		& & AG News & 51.09 & 3e-4\\
		\hline\hline
	\end{tabular}\label{tab3}
\end{table}

\begin{figure}[!t]
	\centering
	\includegraphics[width=\linewidth]{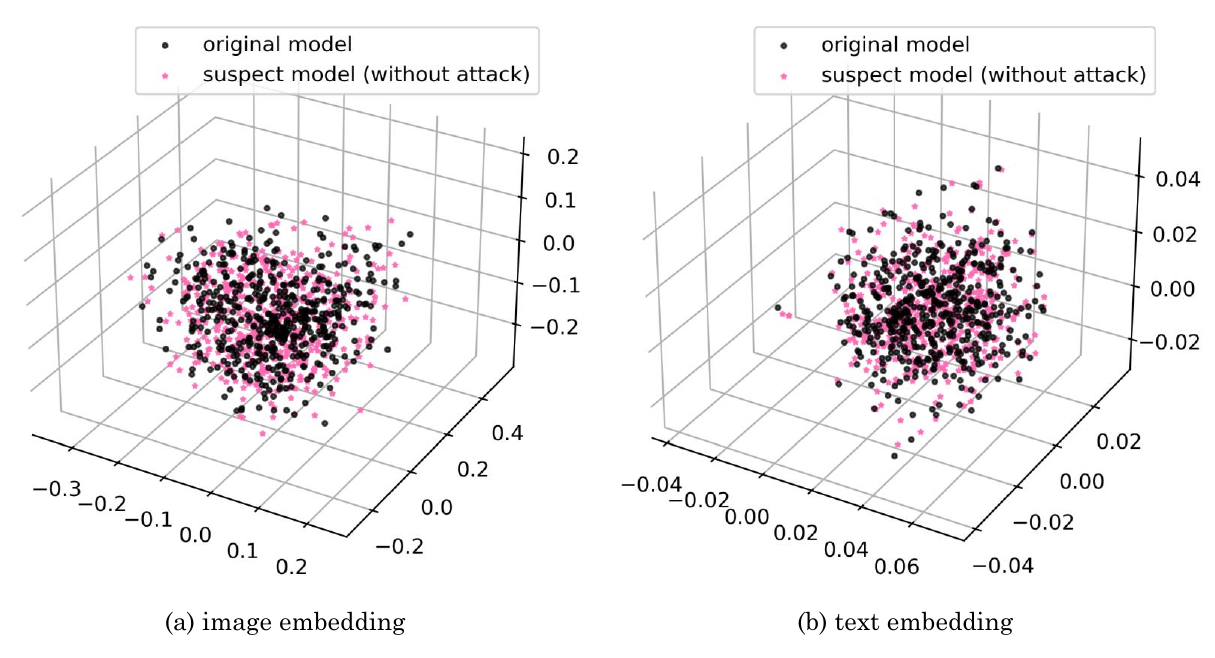}
	\caption{Example for the embedding distribution of the original model and the suspect model in the absence of attacks. The number of embeddings presented is $500\times 2 = 1,000$. The high-dimensional embeddings were reduced to 3-D by randomly selecting three of their components. For image embedding, both the original model and suspect model use Inception-ResNet as the backbone. The same setting applies below unless otherwise stated.
	}\label{NonAttackEmbedding}
\end{figure}

\begin{figure*}[!t]
	\centering
	\includegraphics[width=\linewidth]{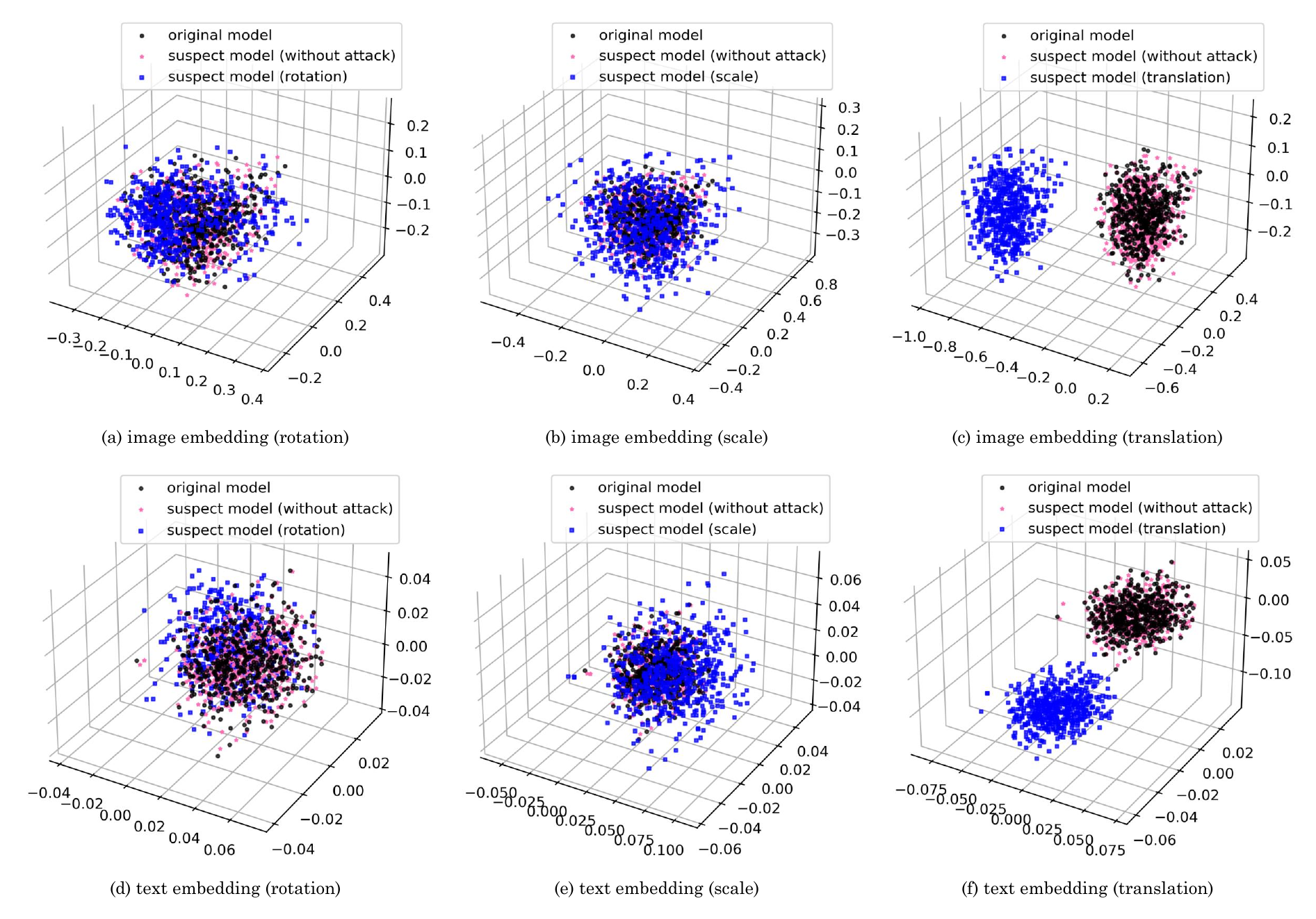}
	\caption{Example for the embedding distribution of the original model and the suspect model under a single attack: (a, d) the rotation angle is $60^\circ$, (b, e) the scaling factor is $2$, (c, f) the translation length is $5$. The rotation direction and translation direction are randomly specified in advance.}\label{SingleAttackEmbedding}
\end{figure*}

\section{Experimental Results and Analysis}
\subsection{Setup}
We conduct extensive experiments on two most representative tasks, i.e., image embedding and text embedding. Evaluating our method on other embedding tasks is an incremental work, which is not the main focus of this paper. For image embedding, we use CASIA-WebFace dataset \cite{Yi:arXiv:2014}, which comprises around $5\times 10^5$ annotated face images spanning around $1\times 10^4$ identities, for model training, and Labeled Faces in the Wild (LFW) \cite{Huang:LFW:2008}, which contains over $1.3\times 10^4$ face images from over $5.5\times 10^3$ unique individuals, for model evaluation. The ratio between the training set and the validation set is 9:1. Models are trained with triplet loss \cite{Schroff:CVPR:2015} to learn discriminative face embeddings. Two backbone architectures, i.e., MobileNet \cite{Howard:arXiv:2017} and Inception-ResNet \cite{Szegedy:AAAI:2017}, are tested as the EaaS model. Verification performance is measured by binary classification accuracy, i.e., to judge whether two face images belong to the same individual. We randomly select 6,000 pairs of faces from LFW and label whether they belong to the same individual. During evaluation, each face in a pair is fed into the encoder, producing two embeddings to be further fed into a simple two-layer fully connected prediction network to determine whether the two faces belong to the same individual.

For text embedding, we use the pretrained model developed by OpenAI, i.e., GPT-3 text-embedding-002\footnote{URL: \url{https://platform.openai.com/docs/models/text-embedding-ada-002}}, for experiments. There are two reasons for using this publicly available embedding model: 1) it is difficult for the authors to train a powerful text embedding model locally due to the large amount of data and computational resources required, and; 2) using a public model better simulates real-world scenarios. The downstream tasks are evaluated using four distinct datasets including Enron Spam \cite{Metsis:CEAS:2006},  SST2 \cite{Socher:EMNLP:2013}, MIND \cite{Wu:ACL:2020} and AG News \cite{Zhang:NIPS:2019}. Enron Spam is a binary-labeled email corpus including over $3.3\times 10^4$ email messages. SST2 is a binary sentiment analysis dataset, including over $6.8\times 10^4$ movie reviews. MIND is a dataset for news recommendation research, containing over $1.6\times 10^5$ records with $20$ categories. AG News  is a news article corpus, including over $1.2\times 10^5$ samples, categorized into four topics, widely used for text classification. The feature dimension for image embedding is $128$, and $1536$ for text embedding. We set $M = 3, N = 5,000$. For image embedding, the $M$ clean models are generated by applying different learning rates to the original model, whereas for text embedding, we generate the models by adding a trainable two-layer fully connected adapter module (each layer contains $1536$ nodes) to the pretrained model. The models are fine-tuned on the `all-nli' dataset\footnote{URL: https://huggingface.co/datasets/sentence-transformers/all-nli} with triplet loss and different learning rates, where only the new module is optimized. Statistical significance is evaluated using the same threshold $p \leq 10^{-3}$, to ensure rigorous and consistent hypothesis testing across all the experimental validations. We will also release source code\footnote{URL: TBD.} to make sure that the reader can easily master the implementation details.

\subsection{Main Results}
We show that RST will not significantly degrade the downstream performance of a model. To verify this argument, for image embedding, we extract face pairs from the test set and judge whether they belong to the identical person. For text embedding, we train the identical prediction network using the embeddings generated by the encoder as input and report the accuracy to assess the utility of the provided embeddings. We apply a mixed attack to embeddings, where the rotation angle is randomly sampled in range $[-180^\circ, 180^\circ]$, the scaling factor is randomly sampled in range $[0.1, 10]$, and each component of the translation vector is randomly sampled in range $[-10, 10]$. It is noted that, we need to randomly specify a uniform rotation or translation direction for the entire point cloud in advance. The sample applies below. All embeddings are processed with the same attack parameters in an experiment. We repeat each experiment ten times, and use the mean as the result. As shown in Table \ref{tab1}, the classification accuracy for attacked embeddings is close to the non-attacked ones in all cases, indicating that the performance can be maintained well. The proposed method does not impair the downstream performance at all since it does not alter the process of model training. Therefore, the most important thing is to verify the fingerprint.

\begin{table*}[!t]
	\centering
	\caption{The fingerprinting performance under the rotation attack scenario. Here, e.g., ``32.51 / 9e-4'' means ``$t$-value / $p$-value''.}
	\begin{tabular}{c|c|c|cccccc}
		\hline\hline
		\multirow{2}{*}{\makecell{Task}} & \multirow{2}{*}{\makecell{Suspect model}} & \multirow{2}{*}{\makecell{Dataset}} & 
		\multicolumn{6}{c}{Rotation degree (random direction specified in advance)}\\
		& & & $-180^\circ$ & $-120^\circ$ & $-60^\circ$ & $30^\circ$ & $90^\circ$ & $150^\circ$ \\
		\hline
		\multirow{2}{*}{\makecell{Image\\embedding}} & MobileNet & \multirow{2}{*}{LFW} & 32.51 / 9e-4 & 32.42 / 9e-4 & 32.76 / 9e-4 & 33.02 / 9e-4 & 31.39 / 1e-3 & 32.00 / 9e-4\\
		\cline{2-2}\cline{4-9}
		& Inception-ResNet & & 37.24 / 7e-4 & 37.64 / 7e-4 & 38.42 / 6e-4 & 38.60 / 6e-4 & 39.60 / 6e-4 & 36.84 / 7e-4 \\
		\hline
		\multirow{4}{*}{\makecell{Text\\embedding}} & \multirow{4}{*}{\makecell{Fully connected\\network with\\skip connection}} & Enron Spam & 39.77 / 6e-4 & 39.04 / 6e-4 & 37.22 / 7e-4 & 36.35 / 7e-4 & 38.28 / 6e-4 & 37.92 / 6e-4 \\
		\cline{3-9}
		& & SST2 & 24.07 / 1e-3 & 23.77 / 1e-3 & 24.37 / 1e-3 & 24.38 / 1e-3 & 24.19 / 1e-3 & 24.03 / 1e-3 \\
		\cline{3-9}
		& & MIND & 158.21 / 3e-5 & 158.69 / 3e-5 & 161.07 / 3e-5 & 147.54 / 4e-5 & 189.01 / 2e-5 & 161.06 / 3e-5 \\
		\cline{3-9}
		& & AG News & 49.99 / 3e-4 & 48.11 / 4e-4 & 51.71 / 3e-4 & 48.13 / 4e-4 & 54.63 / 3e-4 & 54.32 / 3e-4 \\
		\hline\hline
	\end{tabular}\label{tab4}
\end{table*}

\begin{table*}[!t]
	\centering
	\caption{The fingerprinting performance under the scaling attack scenario. Here, e.g., ``33.61 / 8e-4'' means ``$t$-value / $p$-value''.}
	\begin{tabular}{c|c|c|cccccc}
		\hline\hline
		\multirow{2}{*}{\makecell{Task}} & \multirow{2}{*}{\makecell{Suspect model}} & \multirow{2}{*}{\makecell{Dataset}} & 
		\multicolumn{6}{c}{Scaling factor}\\
		& & & $0.2$ & $0.4$ & $0.8$ & $2$ & $4$ & $8$ \\
		\hline
		\multirow{2}{*}{\makecell{Image\\embedding}} & MobileNet & \multirow{2}{*}{LFW} & 33.61 / 8e-4 & 31.86 / 9e-4 & 32.22 / 9e-4  & 33.25 / 9e-4 & 33.42 / 8e-4 & 32.33 / 9e-4\\
		\cline{2-2}\cline{4-9}
		& Inception-ResNet & & 37.02 / 7e-4 & 38.93 / 6e-4 & 39.78 / 6e-4 & 39.12 / 6e-4 & 39.38 / 6e-4 & 37.74 / 7e-4 \\
		\hline
		\multirow{4}{*}{\makecell{Text\\embedding}} & \multirow{4}{*}{\makecell{Fully connected\\network with\\skip connection}} & Enron Spam & 38.35 / 6e-4 & 39.44 / 6e-4 & 38.77 / 6e-4 & 39.83 / 6e-4 & 38.66 / 6e-4 & 40.06 / 6e-4 \\
		\cline{3-9}
		& & SST2 & 24.69 / 1e-3 & 24.10 / 1e-3 & 24.33 / 1e-3 & 23.45 / 1e-3 & 23.83 / 1e-3 & 24.39 / 1e-3 \\
		\cline{3-9}
		& & MIND & 168.87 / 3e-5 & 150.36 / 4e-5 & 179.97 / 3e-5 & 153.22 / 4e-5 & 161.82 / 3e-5 & 147.42 / 4e-5 \\
		\cline{3-9}
		& & AG News & 50.52 / 3e-4 & 53.19 / 3e-4 & 49.99 / 3e-4 & 52.80 / 3e-4 & 52.43 / 3e-4 & 49.77 / 4e-4 \\
		\hline\hline
	\end{tabular}\label{tab5}
\end{table*}

\begin{table*}[!t]
	\centering
	\caption{The fingerprinting performance for under the translation attack scenario. Here, e.g., ``34.50 / 8e-4'' means ``$t$-value / $p$-value''.}
	\begin{tabular}{c|c|c|cccccc}
		\hline\hline
		\multirow{2}{*}{\makecell{Task}} & \multirow{2}{*}{\makecell{Suspect model}} & \multirow{2}{*}{\makecell{Dataset}} & 
		\multicolumn{6}{c}{Translation length (random direction specified in advance)}\\
		& & & $1$ & $2$ & $4$ & $6$ & $8$ & $10$ \\
		\hline
		\multirow{2}{*}{\makecell{Image\\embedding}} & MobileNet & \multirow{2}{*}{LFW} & 34.50 / 8e-4 & 32.84 / 9e-4 & 33.60 / 8e-4 & 31.58 / 1e-3 & 33.15 / 9e-4 & 31.89 / 1e-3\\
		\cline{2-2}\cline{4-9}
		& Inception-ResNet & & 38.51 / 6e-4 & 38.41 / 6e-4 & 38.43 / 6e-4 & 37.01 / 7e-4 & 37.36 / 7e-4 & 40.45 / 6e-4 \\
		\hline
		\multirow{4}{*}{\makecell{Text\\embedding}} & \multirow{4}{*}{\makecell{Fully connected\\network with\\skip connection}} & Enron Spam & 38.78 / 6e-4 & 38.45 / 6e-4 & 37.39 / 7e-4 & 39.77 / 6e-4 & 37.93 / 6e-4 & 38.70 / 6e-4 \\
		\cline{3-9}
		& & SST2 & 24.05 / 1e-3 & 24.38 / 1e-3 & 23.41 / 1e-3 & 23.16 / 1e-3 & 23.51 / 1e-3 & 23.40 / 1e-3 \\
		\cline{3-9}
		& & MIND & 154.21 / 4e-5 & 146.42 / 4e-5 & 179.20 / 3e-5 & 168.21 / 3e-5 & 158.31 / 3e-5 & 161.11 / 3e-5 \\
		\cline{3-9}
		& & AG News & 52.62 / 3e-4 & 52.18 / 3e-4 & 52.10 / 3e-4 & 51.75 / 3e-4 & 48.06 / 4e-4 & 50.47 / 3e-4 \\
		\hline\hline
	\end{tabular}\label{tab6}
\end{table*}

\begin{figure}[!t]
	\centering
	\includegraphics[width=\linewidth]{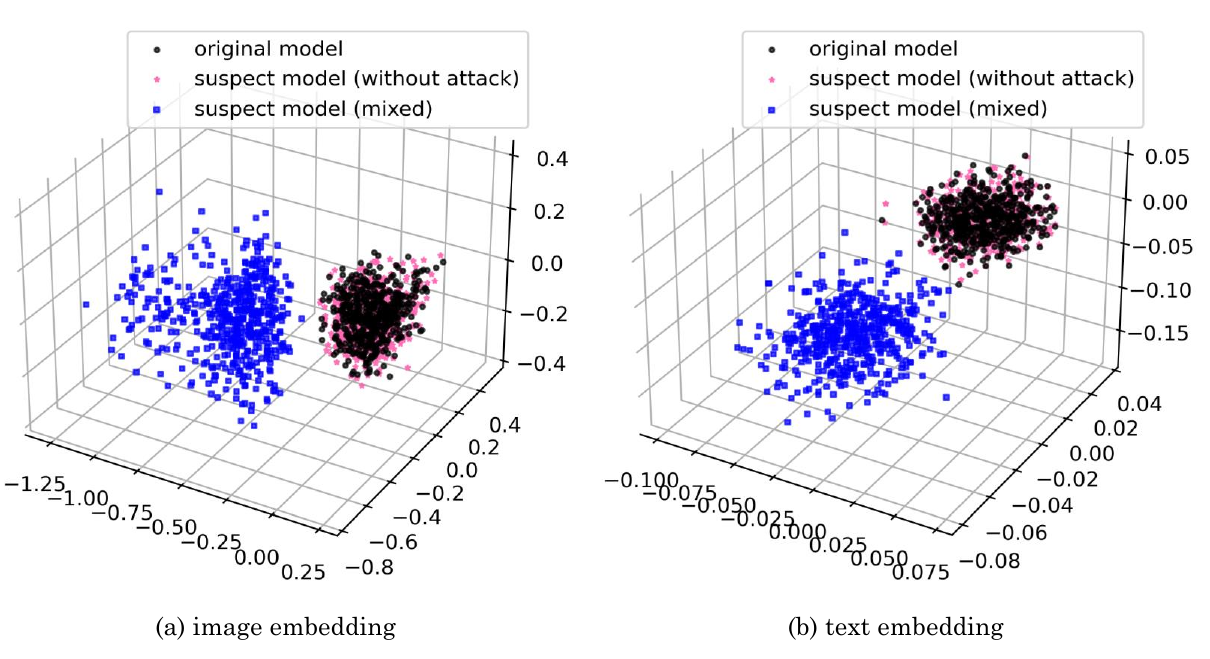}
	\caption{Example for the embedding distribution of the original model and the suspect model under a combined attack with the parameters given in Fig. \ref{SingleAttackEmbedding}. }\label{MixedAttackEmbedding}
\end{figure}

\begin{table}[!t]
	\centering
	\caption{The fingerprinting performance under the mixed attack scenario.}
	\begin{tabular}{c|c|c|c|c}
		\hline\hline
		Task & Suspect model & Dataset & $t$-value & $p$-value\\
		
		\hline
		
		\multirow{2}{*}{\makecell{Image\\embedding}} & MobileNet & \multirow{2}{*}{LFW} & 32.06 & 9e-4\\
		
		\cline{2-2}\cline{4-5}
		& Inception-ResNet & & 39.22 & 6e-4\\
		
		\hline
		
		\multirow{4}{*}{\makecell{Text\\embedding}} & \multirow{4}{*}{\makecell{Fully connected\\network with\\skip connection}} & Enron Spam & 37.42 & 7e-4\\
		
		\cline{3-5}
		& & SST2 & 24.41 & 1e-3\\
		
		\cline{3-5}
		& & MIND & 192.83 & 2e-5\\
		
		\cline{3-5}
		& & AG News & 51.72 & 3e-4\\
		\hline\hline
	\end{tabular}\label{tab7}
\end{table}

We first describe the architecture of the suspect model used in our experiments. For image embedding, the suspect model uses the same backbone architecture as the victim model. For text embedding, the suspect model includes six fully connected layers each of which contains $1536$ nodes. Moreover, there is a skip connection between the first layer and the fourth layer. Table \ref{tab2} reports the classification accuracy on the downstream tasks of suspect model. It is inferred from Table \ref{tab1} and \ref{tab2} that, regardless of whether the suspected model differs significantly or slightly in structure from the original model, we are able to effectively steal the model's functionality, which indicates that we can replicate the downstream performance of the model. 

Fig. \ref{NonAttackEmbedding} shows the distribution of embedding vectors produced by the original model and by the suspect model in the absence of attacks, given the same inputs. It is observed that, intuitively, the distributions of the two sets of embedding vectors are very similar. Table \ref{tab3} reports the fingerprinting performance for this non-attack scenario. It can be found that the $p$-values are all no more than $10^{-3}$, indicating that, the model fingerprint can be reliably extracted and verified. Below, we analyze the fingerprint verification performance after attack.

Since RST transformations involve different transformation parameters, we should analyze the impact of individual attacks on fingerprinting. Fig. \ref{SingleAttackEmbedding} demonstrates concrete examples for the distribution of embedding vectors produced by the original model and by the suspect model under a single attack. Despite visual differences in the distributions, they still exhibit a high degree of similarity, which is in line with the theoretical results we presented earlier. Table \ref{tab4}, Table \ref{tab5} and Table \ref{tab6} show the fingerprinting performance after applying the rotation attack, scaling attack, and transformation attack, respectively, due to different attack parameters. It is observed that all $p$-values fall within the verifiable range, which has verified the robustness.

As shown in Fig. \ref{MixedAttackEmbedding}, the attacker may apply a combined attack (or say mixed attack) to the embedding vectors generated by the suspect model, which can cause significant distribution difference between different sets of embedding vectors. It is necessary to evaluate the fingerprint verification performance after a mixed attack. To this end, we simulate the mixed attack in experiments, where the rotation angle is randomly sampled in range $[-180^\circ, 180^\circ]$, the scaling factor is randomly sampled in range $[0.1, 10]$, and each component of the translation vector is randomly sampled in range $[-10, 10]$. All embeddings are processed with the same attack parameters in an experiment. We repeat the experiment $10$ times, and use the mean $t$-value as the result. It can be inferred from the experimental results shown in Table \ref{tab7} that the proposed method still successfully verifies the model's fingerprint, thereby once again verifying the superiority, reliability, and practicality of our method.

\begin{table*}[!t]
	\centering
	\scriptsize
	\caption{The fingerprinting performance due to different data sizes under the mixed attack scenario.}
	\begin{tabular}{c|c|c|cccccc}
		\hline\hline
		\multirow{2}{*}{\makecell{Task}} & \multirow{2}{*}{\makecell{Suspect model}} & \multirow{2}{*}{\makecell{Dataset}} & 
		\multicolumn{6}{c}{Data size (i.e., the total number of data samples used for fingerprint verification)}\\
		& & & $100$ & $200$ & $500$ & $1000$ & $2000$ & $5000$ \\
		\hline
		\multirow{2}{*}{\makecell{Image\\embedding}} & MobileNet & \multirow{2}{*}{LFW} & -11.08 / 8e-3 & 17.20 / 3e-3 & 29.43 / 1e-3 & 34.86 / 8e-4 & 33.06 / 9e-4 & 32.06 / 9e-4 \\
		\cline{2-2}\cline{4-9}
		& Inception-ResNet & & -6.39 / 2e-2 & 17.95 / 3e-3 & 27.33 / 1e-3 & 31.24 / 1e-3 & 32.44 / 9e-4 & 39.22 / 6e-4 \\
		\hline
		\multirow{4}{*}{\makecell{Text\\embedding}} & \multirow{4}{*}{\makecell{Fully connected\\network with\\skip connection}} & Enron Spam & -2.08 / 1e-1 & -1.30 / 3e-1 & -0.53 / 6e-1 & 1.66 / 2e-3 & 38.72 / 6e-4 & 37.42 / 7e-4 \\
		\cline{3-9}
		& & SST2 & -1.85 / 2e-1 & 0.84 / 4e-1 & 0.54 / 6e-1 & 3.91 / 5e-2 & 23.88 / 1e-3 & 24.41 / 1e-3 \\
		\cline{3-9}
		& & MIND & -4.11 / 5e-2 & -3.10 / 9e-2 & 1.94 / 1e-1 & 19.51 / 2e-3 & 169.11 / 3e-5 & 192.83 / 2e-5 \\
		\cline{3-9}
		& & AG News & -2.54 / 1e-1 & -1.45 / 2e-1 & -0.57 / 6e-1 & 1.24 / 3e-1 & 49.54 / 4e-4 & 51.72 / 3e-4 \\
		\hline\hline
	\end{tabular}\label{tab8}
\end{table*}

\begin{table*}[!t]
	\centering
	\scriptsize
	\caption{The similarity between the embeddings produced by the original model and that produced by the suspect model under different scenarios. E.g., ``329.24'' means the similarity between the embeddings produced by the original model and that produced by the suspect model under the mixed attack scenario. The results reported here are mean values.}
	\begin{tabular}{c|c|c|c|c|c|c}
		\hline\hline
		\makecell{Task} & \makecell{Suspect model} & \makecell{Dataset} & original model & suspect model (without attack) & suspect model (mixed attack) & suspect model (mixed, aligned)\\
		\hline
		\multirow{2}{*}{\makecell{Image\\embedding}} & MobileNet & \multirow{2}{*}{LFW} & 0 & 0.29 & 329.24 & 0.29 \\
		\cline{2-2}\cline{4-7}
		& Inception-ResNet & & 0 & 0.26 & 25.65 & 0.25 \\
		\hline
		\multirow{4}{*}{\makecell{Text\\embedding}} & \multirow{4}{*}{\makecell{Fully connected\\network with\\skip connection}} & Enron Spam & 0 & 0.04 & 25.36 & 0.03 \\
		\cline{3-7}
		& & SST2 & 0 & 0.03 & 25.38 & 0.02 \\
		\cline{3-7}
		& & MIND & 0 & 0.02 & 25.26 & 0.02 \\
		\cline{3-7}
		& & AG News & 0 & 0.03 & 25.18 & 0.02 \\
		\hline\hline
	\end{tabular}\label{tab9}
\end{table*}

\begin{figure}[!t]
	\centering
	\includegraphics[width=\linewidth]{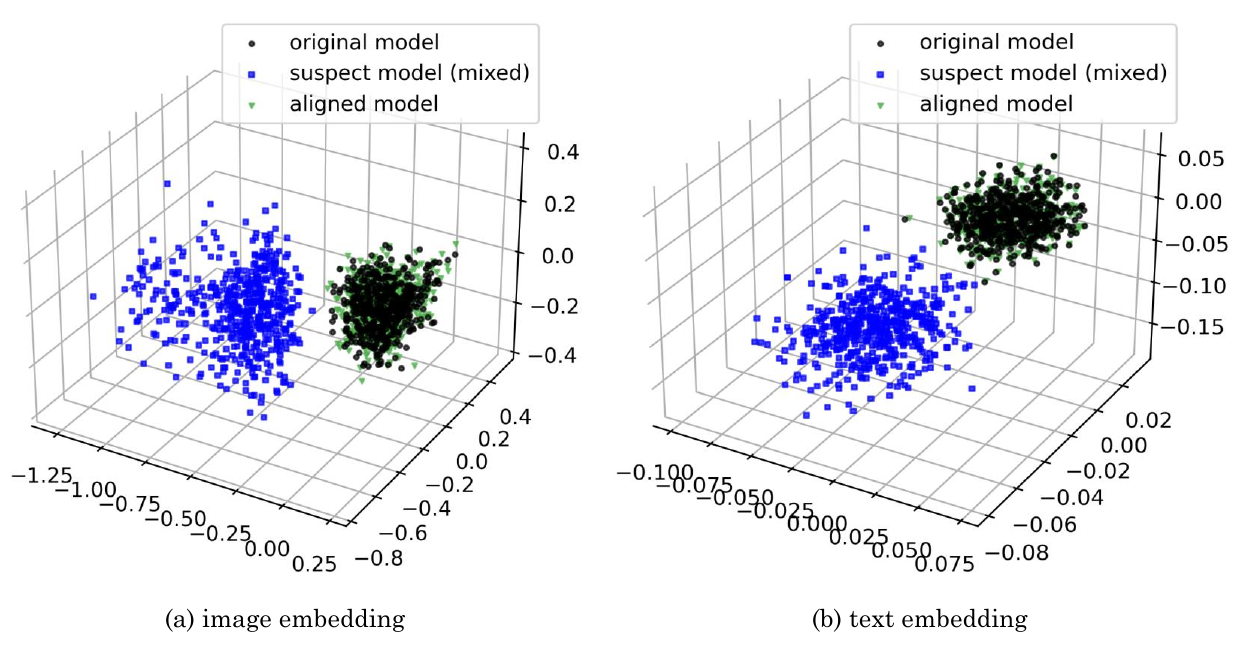}
	\caption{The aligned embedding distribution for the example presented in Fig. \ref{MixedAttackEmbedding}. Here, ``aligned model'' means processing the attacked embeddings using the estimated attack parameters to align them with the distribution of embeddings produced by the original model. }\label{AlignedMixedAttackEmbedding}
\end{figure}

\subsection{Ablation Study and Other Analysis}
In the proposed method, we need to specify the total number of sample points used for fingerprint verification in advance. To analyze the effect of sample size, we conduct experiments with the number of points ranging from $100$ to $5,000$ under the mixed attack scenario (i.e., the most complex attack scenario). It is inferred from Table \ref{tab8} that different data sizes generally result in different fingerprinting performance. From the overall trend, as the data size increases, the success rate of fingerprint verification becomes higher, which is due to that a larger data size will enhance statistical significance. In our experiments, we set the data size to $5,000$, which is sufficient for fingerprint verification, as shown in Table \ref{tab8}.

In fact, since the proposed method requires the estimation of attack parameters, it is quite desirable to evaluate the reliability of the parameter estimation, which can be used to validate and support the correctness of the previous experimental results. We consider the most common scenario, i.e., mixed attack. Fig. \ref{AlignedMixedAttackEmbedding} presents the aligned embedding distribution for the example given in Fig. \ref{MixedAttackEmbedding}. It can be observed that the aligned embeddings cluster together with that produced by the original model. This indicates that the proposed alignment method is effective. Eq. (32) can be slightly altered to compute the similarity between any two point clouds. We here use it to compute the similarity between the embeddings produced by the original model and that produced by the suspect model under different scenarios. Experimental results are given in Table \ref{tab9}, which uses the same settings as Table \ref{tab7}. It can be seen that the similarity between the original embeddings and the aligned embeddings is close to the similarity between the original embeddings and the embeddings produced by the suspect model without attack, implying that the proposed parameter estimation method well aligns the attacked embeddings with the original embeddings, thereby supporting robust EaaS model fingerprint verification.

The proposed POSTER method requires estimating different attack parameters. It is necessary to analyze the time complexity of estimating these parameters. Specifically, although the estimation of the scaling factor requires us to solve the least squares problem described in Eq. (29), its closed-form solution given in Eq. (30) allows us to quickly find the answer. At this point, the main time consumption lies in matrix multiplication and the computation of Moore-Penrose inverse, both of which can be efficiently calculated by parallel computing and existing techniques. For estimating the rotation matrix and translation vector, we need to compute the centroid, centralize the point cloud, as well as calculate the covariance matrix. These routine operations are not computationally intensive. For example, the calculations of different elements in the covariance matrix can be independent of each other, allowing for parallel computing. The main time consumption lies in SVD and matrix multiplication. The former is proportional to the embedding dimension, while the latter is proportional to the number of cloud points. By controlling the embedding dimension, the number of cloud points, and using powerful computers, the time complexity can be kept at an acceptable level. These requirements can be met in real commercial scenarios, demonstrating the applicability.

\section{Conclusion and Discussion}
In this paper, we propose, for the first time, a fingerprinting method for EaaS models that is robust against rotation, scaling, and translation attacks - typical threats in real-world scenarios. The proposed method does not participate in model training and can achieve robust fingerprint verification under black-box scenarios by leveraging the spatial distribution characteristics of feature point clouds, while also resisting the aforementioned RST attacks. Its technical rationale lies in the fact that the distribution characteristics of point clouds are inherent features produced by the model during training, which can be regarded as a robust fingerprint of the model. In cases of intellectual property disputes or when verifying the origin of an encoder model, fingerprint verification can be performed by computing and examining the consistency of the point cloud distributions. Our experimental results have shown that the proposed method enables the fingerprint to be robustly verified while preserving the performance of the embedding points on downstream tasks, demonstrating superior performance resistance to RST attacks. In future, we will improve the proposed method to resist more real-world attacks and hope this primary attempt can contribute to the community and inspire more advanced works. 

\section*{Acknowledgment}
This work was partly supported by National Natural Science Foundation of China under Grant Number U23B2023, Science and Technology Commission of Shanghai Municipality under Grant Number 24ZR1424000, and the Basic Research Program for Natural Science of Guizhou Province under Grant Number QIANKEHEJICHU-ZD[2025]-043.

\end{document}